# Molecular Gas Clumps from the Destruction of Icy Bodies in the β Pictoris Debris Disk

*Science (2014) 343, 1490*

W.R.F. Dent[1*], M.C. Wyatt[2], A. Roberge[3], J.-C. Augereau[4], S. Casassus[5], S. Corder[1], J.S. Greaves[6], I. de Gregorio-Monsalvo[1], A. Hales[1], A.P. Jackson[2], A. Meredith Hughes[7], A.-M. Lagrange[4], B. Matthews[8], D. Wilner[9]

1) ALMA Santiago Central Offices, Alonso de Córdova 3107, Vitacura, Casilla 763 0355, Santiago, Chile.

2) Institute of Astronomy, Madingley Road, Cambridge CB3 0HA, UK.

3) Exoplanets and Stellar Astrophysics Lab, NASA Goddard Space Flight Center, Greenbelt, MD, 20771, USA.

4) UJF-Grenoble 1 / CNRS-INSU, Institut de Planétologie et d'Astrophysique de Grenoble (IPAG), UMR 5274, Grenoble, F-38041, France.

5) Departamento de Astronomía, Universidad de Chile, Casilla 36-D, Santiago, Chile.

6) Dept. of Astronomy, University of St. Andrews, North Haugh, St. Andrews, UK.

7) Wesleyan University Department of Astronomy, Van Vleck Observatory, 96 Foss Hill Drive, Middletown, CT 06459, USA.

8) National Research Council of Canada, Herzberg Astronomy & Astrophysics Programs, 5701 West Saanich Road, Victoria, BC, Canada, V9E 2E7, and Department of Physics & Astronomy, University of Victoria, Finnerty Road, Victoria, BC, V8P 5C2, Canada

9) Smithsonian Astrophysical Observatory, 60 Garden St., MS 42, Cambridge, MA 02138 USA.

* Correspondence to wdent@alma.cl

**Abstract:**

Many stars are surrounded by disks of dusty debris formed in the collisions of asteroids, comets and dwarf planets. But is gas also released in such events? Observations at submm wavelengths of the archetypal debris disk around β Pictoris show that 0.3% of a Moon mass of carbon monoxide orbits in its debris belt. The gas distribution is highly asymmetric, with 30% found in a single clump 85AU from the star, in a plane closely aligned with the orbit of the inner planet, β Pic b. This gas clump delineates a region of enhanced collisions, either from a mean motion resonance with an unseen giant planet, or from the remnants of a collision of Mars-mass planets.

**Main Text:**

Debris disks are the end product of collisional cascades of km-sized bodies orbiting stars (including the Sun), and are normally thought to contain negligible gas. At a distance of 19.44 parsec (*1*) and age of 20 million years (*2*), β Pictoris is one of the closest, brightest and youngest examples. Its edge-on disk was the first to be imaged in scattered light, showing the distribution of micron-sized dust grains (*3*). Various subsequent observations have provided evidence of infalling comets within a few Astronomical Units (AU) of the star (*4*), a massive planet at ~10 AU (*5*), as well as atomic gas extending out to ~300AU (*6*); it is still unclear how these features are linked. By observing such debris disks at mm wavelengths, it is possible to trace the mm-sized dust and, by inference, the parent bodies of the collisional cascade, known as planetesimals (*7*).

We observed β Pic using the Atacama Large Millimeter/submillimeter Array (ALMA) at a wavelength of 870μm in both the continuum and the J=3-2 $^{12}$CO line with a resolution of 12AU (*8*). The continuum image (Fig.1A) shows maxima in the surface brightness ~60AU either side of the star. At separations from 30-80AU the continuum is, on average, 15% brighter to the southwest. These results indicate that the mm grains lie in a broad slightly asymmetric belt nearly co-located with the disk of sub-micron reflecting dust (*9*). The total flux of 60±6mJy corresponds to a dust mass of 4.7±0.5x10$^{23}$kg (6.4M$_{Moon}$), if we assume a standard dust mass opacity (0.15m$^2$ kg$^{-1}$ at 850μm) and temperature of 85K (*10*).

The planet β Pic b was close to maximum SW elongation at the time of the observations, with a projected separation of 8.7AU (*11*). Its location is coincident with a dip in the continuum surface brightness at a significance level of 4σ, suggesting that it may influence the innermost dust distribution.

ALMA also detected the disk in the $^{12}$CO J=3-2 transition (Fig.1B), with a clear velocity gradient along the major axis, illustrated in the position-velocity (PV) diagram (Fig.2). This shows the characteristic distribution of a broad belt of orbiting gas, with inner and outer radii of 50 and 160AU, and a peak around 85AU. No gas emission is seen inside 50AU. The CO distribution is, on average, a factor of 2 brighter to the southwest, and a similar asymmetry was seen in the mid-infrared emission (*12*). Unlike the sub-mm continuum, the CO clump is offset by ~5AU above the main disk plane (Fig.1B) and is more closely aligned with the inner planet β Pic b and a secondary disk seen in scattered light (*5,9*).

The CO J=3-2 emission totals $7.6\pm0.8\times10^{-20}$ W/m$^2$. Assuming a range of possible excitation temperatures from 20K (measured from UV absorption lines (*13*)), to 85K (the dust equilibrium temperature (*10*)), this corresponds to a CO mass of $1.7\times10^{20}$kg (0.0023$M_{Moon}$) within a factor of 2. Towards the star itself, the total CO column is $2.5_{(+2.5,-1.2)}\times10^{15}$ cm$^{-2}$ ; results from UV absorption line measurements range between $0.6-2.1\times10^{15}$ cm$^{-2}$ (*13*), indicating that at least 10% of the CO along this line-of-sight lies in front of the star.

The attenuation through the disk due to dust or CO self-shielding is low and, at radii >50AU, CO is destroyed mostly by UV photons from the ambient interstellar medium (ISM) (*14*). The CO photodissociation timescale in the unshielded outer disk will be ~120yrs (*15*) - substantially less than the 600-year orbital period at 85AU. Unless we are observing β Pic at a very unusual time (i.e. ≤100 years after a collision) then the CO must be continuously replenished, with a steady-state production rate of ~$1.4\times10^{18}$ kg/yr. The source of CO is likely the icy debris (grains, comets and planetesimals) in the disk (*16*). In our own solar system, comets are mostly composed of refractory silicate grains together with $H_2O$ and CO ices, with a typical CO/$H_2O$ ratio of 0.01-0.1 (*17*). Although CO sublimates at ~20K - substantially less than the 85K dust temperature in the β Pic disk - it can be trapped in $H_2O$ ice at temperatures up to ~140K (*18*), which could occur in radiative equlibrium as close as 30AU from β Pic (*14*). At larger distances, photodesorption or collisions can release this CO into the gas phase. Photodesorption (a.k.a. UV sputtering) of mm-sized ice grains (*19*) in a dust clump with 10% of the total disk mass would produce CO at a rate that is ~10 times lower than our observations show. The CO linewidth measured directly towards the star is ~1km/s (Fig.2), suggesting the velocity dispersion is lower than the 6km/s thought necessary for collisions to directly vaporise ice (*20*). However icy parent bodies being shattered in multiple lower velocity collisions may release much of the entrapped CO as gas, leaving the less volatile $H_2O$ ice behind. To sustain the gas production rate with a CO ice release fraction of 0.1, the required mass loss of colliding grains would be ~$1.4\times10^{19}$ kg/yr, equivalent to the destruction of a large comet every 5 minutes. If sustained over the lifetime of β Pic, ~50$M_{Earth}$ of icy bodies will have been removed.

Continual replenishment of CO may explain another puzzle. In the absence of a braking mechanism many of the atomic species such as Na found around β Pic should be blown out from the central star under the force of radiation pressure (*21,22,23*). Carbon is overabundant relative to other metals, by factors of 18 to 400 (*21,22*), and has been proposed as the mechanism that allows the ionised gas to remain bound to the star. C+ couples to other ions through Coulomb forces; and since it feels weak radiation pressure from the central star, it tends to brake all the gas. The origin of the unusual carbon overabundance may be explained if CO is being rapidly released in the clump, then dissociated and ionized before being spread throughout the disk. In steady state, the relative C+/CO abundance implies that C+ is being removed on a timescale of $3\times10^3$-$10^4$yrs; since this is several orbital timescales at 85AU, the braking gas can be spread throughout the disk.

If the gas is on circular Keplerian orbits, each point in the PV diagram corresponds to one of two points in the disk. Fig.3 shows possible deprojections of the data to form the face-on gas distribution (*8*). A compact clump of CO is found at 85AU radius which contains ~30% of the flux, along with an extended 'tail' of emission. The radial distribution of the mm dust can also be derived (*8,12*), and shows excess around this radius in the southwest (Fig.3C). The mm dust shows an abrupt drop in density around 130AU radius, coincident with a change in slope in the radial distribution of scattered light. This was interpreted as a transition from a dust-producing planetesimal belt to a region dominated by submicron-sized grains blown out by radiation pressure (*9,24*).

What is the origin of the CO clumps and corresponding mid-infrared and continuum features? There are two possible interpretations, both requiring planet-mass objects in the outer disk, but with different predictions for the deprojected gas distribution. In the first (Fig.3A), outward migration of a planet traps planetesimals into both the 2:1 and 3:2 mean motion resonances, resulting in a distribution with two clumps of asymmetric brightness on opposite sides of the star that orbit with the planet (*25*). Planetesimal collisions would occur most frequently in the clumps which would then be the most vigorous production sites of both CO and the short-lived micron-sized grains seen in mid-infrared images (*12*). The morphology of the clumps constrains the planet's mass and migration parameters (*25*). Notably the planet must be >10$M_{Earth}$ to have captured material into its 2:1 resonance, and it would currently be close to the inner edge of the gas/dust belt, ~90degrees in front of the SW clump. The CO clump's 'tail' would point in the opposite direction to the rotation, and its' length given by the ratio of CO photodissociation lifetime to the synodic period.

The alternative interpretation (Fig.3B) is that the CO and micron-sized dust originate in a single recent collision (*12,26*). Since the clump also contributes ~10% of the flux in our sub-mm continuum image, the parent body must have ~Mars mass, as giant impacts typically release ~10% of the progenitor's mass as debris (*27*). While collisional debris persists in a clump for ~100yr after such an event, it is more likely that the collision occurred ~0.5Myr ago, with the asymmetry arising because the orbits of all collisional debris pass through the collision point, a region which would retain a high density and collision rate (*26)*. Such a clump would be stationary, and the CO 'tail' would lie in the rotation direction, with a length given by the ratio of photodissociation time to sidereal period (Fig.3B).

Tentative evidence has been published for the clump's motion (*28*), which if confirmed would favour the resonance interpretation. Either way these observations provide a valuable opportunity to ascertain the prevalence of planet-sized objects at large orbital distances in this disk.

The CO gas detected in β Pic is unusual among debris disks. So far, two others have been detected in CO (49 Cet and HD21997 (*16*)) and one in OI (HD172555 (*29*)). Most debris disks are presumed to contain icy grains, so we would expect all such systems to host CO and its photodissociation products. The high collision rates found in clumps may substantially enhance the gas abundance. If CO is formed in collisions, its brightness will scale as the collision rate, which is proportional to $v_r n_d^2$, where $v_r$ is the collision velocity and $n_d$ is the number of dust grains. Both $n_d$ and $v_r$ are enhanced in clumps due to resonances and giant impacts, increasing collision rates by factors of up to 10-100 (*7,26*). The presence of these clumps can increase the overall collision rate and CO flux from a disk by an order of magnitude. The CO and compact clump in the β Pic disk indicates a period of intense activity, driven by planets or planet collisions.

## Acknowledgements:

This paper makes use of the following ALMA data: ADS/JAO.ALMA#2011.0.00087.S. ALMA is a partnership of ESO (representing its member states), NSF (USA) and NINS (Japan), together with NRC (Canada) and NSC and ASIAA (Taiwan), in cooperation with the Republic of Chile. The Joint ALMA Observatory is operated by ESO, AUI/NRAO and NAOJ. Partial financial support for SC, MCW and AH was provided by Millenium Nucleus P10-022-F (Chilean Ministry of Economy). AR acknowledges support by the Goddard Center for Astrobiology, part of the NASA Astrobiology Institute. MCW was supported by the European Union through ERC grant number 279973, IdG acknowledges support from MICINN (Spain) AYA2011-30228-C03 grant (including FEDER funds), APJ was supported by an STFC postgraduate studentship, and JCA acknowledges the French National Research Agency (ANR) for support through contract ANR-2010 BLAN-0505-01 (EXOZODI).

**Figures**

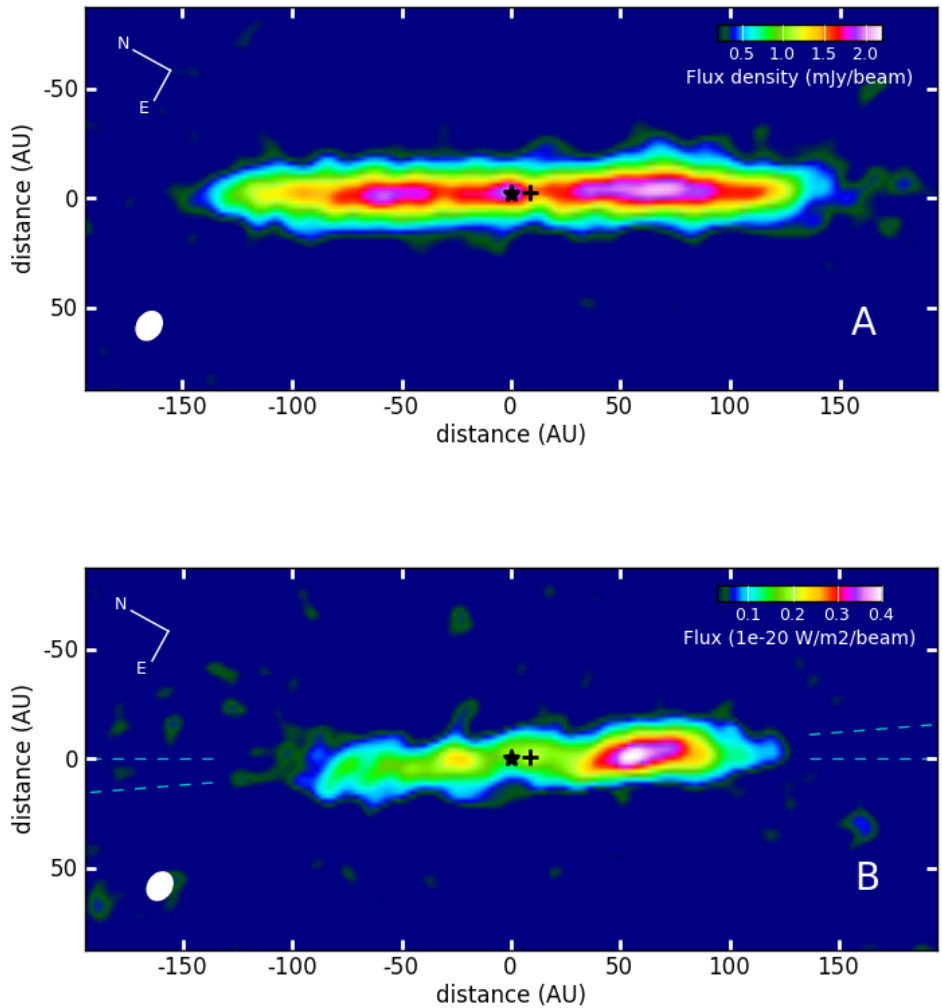

**Fig.1.** Images of β Pic using ALMA in continuum and line emission. Both have been rotated by +29 degrees, the position angle of the main dust disk. The beam size (0.56x0.71arcsec) is shown lower left, and the locations of the star and planet on the date of observations are indicated by the asterisk and the cross. (A) Continuum emission at 870μm from the mm dust; the rms noise level is 61μJy beam$^{-1}$. (B) Total J=3-2 CO line emission (rest frequency 345.796GHz); the noise level is 0.02x10$^{-20}$ Wm$^{-2}$. The planes of the main and secondary dust disks (*9*) are shown by the dashed lines.

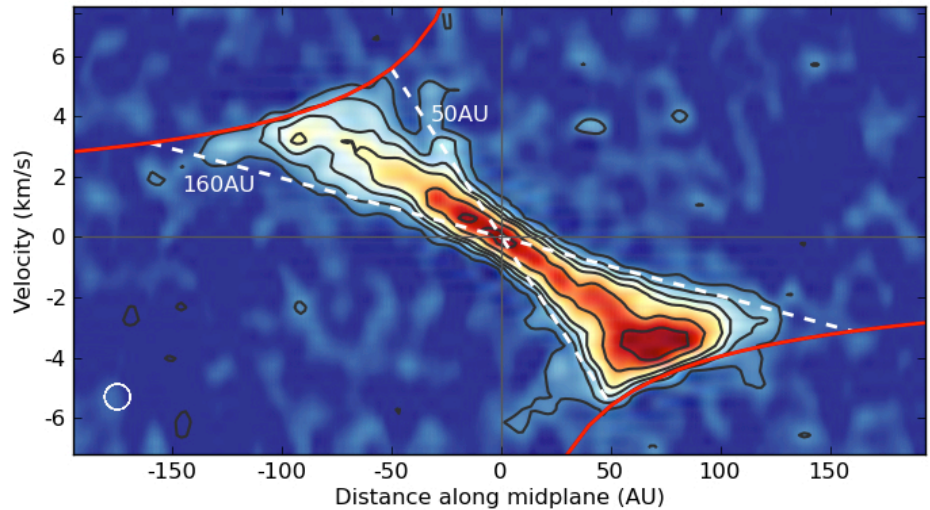

**Fig. 2.** Velocity distribution along disk major axis. This shows the observed velocity of the CO emission relative to the velocity measured towards the star (which has a centroid of +20.3±0.03 km/s, in the barycentric reference frame). Gas on the brighter SW side is approaching us; the two dashed white lines represent the velocity we would measure from gas at the inner and outer radii of the CO belt. The solid red lines show the true orbital velocity, the maximum observed velocity from gas at the tangential point, as a function of distance from the star, assuming a stellar mass of 1.75$M_{solar}$. Contours are 10, 20, 30, 40, 60, 80% of the peak (0.06 Jy/beam). The spectral/spatial resolution is illustrated lower left.

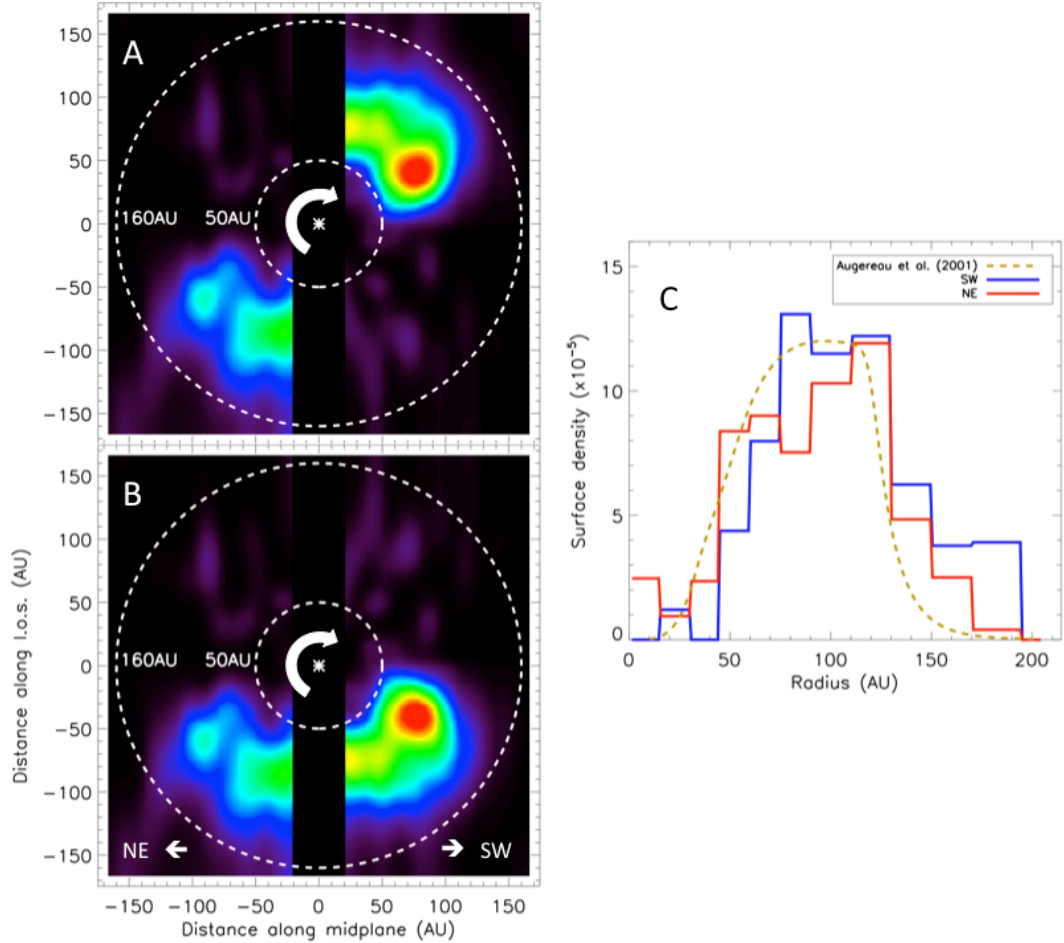

**Fig. 3.** Deprojected distributions of gas and dust around β Pic (*8*). (A,B) Two possible face-on distributions of the CO assuming circular orbital motion. (A) The two-clump distribution, interpreted as mean-motion resonances with an inner planet. In this case the bright southwest resonance lies on the far side of the star and is approaching us. (B) Single clump distribution, interpreted as a collision of ~Mars-mass objects; in this case the southwest clump is stationary and the 'tail' of CO points in the direction of rotation. (C) Radial distribution of mm-sized dust in the southwest and northeast sectors of the disk, obtained by fitting consecutive annuli to the continuum distribution on the two sides (*12*). The dashed line is an axisymmetric model for the parent body distribution predicted from scattered light images of the sub-micron grains (*24*). Surface density units are cross-section area per unit disk area, in $AU^2/AU^2$.

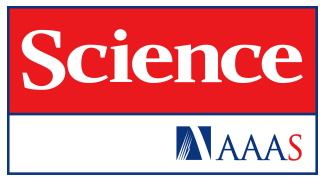

# Supplementary Materials

Methods

The observations at a wavelength around 870μm were performed on Nov 16, 2012, using ALMA with 27 antennas with projected baseline lengths from 15 to 380m. This resulted in a synthesized beam of 0.56x0.71arcsec (FWHM) at PA=62deg. At the distance of β Pic, these interferometric observations were sensitive to emission on scales from 10 to 240AU. The absolute astrometric accuracy of the sub-mm images is better than 0.1arcsec. Two field centres were observed, separated by +/- 5 arcsec along the disk major axis; with the primary beam of 18 arcsec (FWHM), this allowed us to cover the disk with more even sensitivity. The images in Fig.1 have been corrected for the response of the primary beam.

Callisto was used as an amplitude calibrator, with J0522-364 to calibrate the bandpass, and J0519-4546 as the nearby phase/gain calibrator. Including calibrators and overheads, the total observing time was 2.7hrs, with 1hr integration on the source. The flux calibration accuracy from such ALMA data is estimated to be 10%.

The data were reduced using CASA and imaged using the clean algorithm, with natural weighting of the antenna uv response; the final images were produced with a 0.1 arcsec pixel spacing.

The correlator was set to observe simultaneously four 2GHz-wide spectral windows, which were combined to make the continuum data, with a total bandwidth of 7.4GHz centred at 340GHz, after removal of the channels with line emission. The rms sensitivity of the continuum map was 61μJy/beam, giving a signal:noise of 33 at the peak position. The $^{12}$CO J=3-2 line at a rest frequency of 345.796GHz was observed in one of the spectral windows, with a channel spacing of 0.488MHz, and a spectral resolution of 0.85km/s (after Hanning smoothing). The resultant datacube had an rms of 0.004Jy/channel/beam, giving a maximum signal:noise of 25 in the brightest spectral channel in the disk.

Interpretation of Position-velocity diagrams of a Keplerian non-axisymmetric disk

To obtain the Position-velocity diagram in Fig.2, a cut was taken along the disk major axis and the emission was integrated over ±1 arcsec orthogonal to the disk plane. While the disk in β Pic is almost edge-on, the distance of the gas along the line of sight can be determined using the velocity information, assuming the gas to be on circular orbits at the Keplerian velocity of a 1.75$M_{sun}$ star. Thus it is possible to deproject the Position-velocity diagram of Fig. 2, and to place all of the emission in that figure into another showing the face-on distribution of gas (see Fig. 3A,B). However, there are two solutions for the gas location, on the near and far side of the disk with respect to the central star. For the deprojections shown in Fig. 3, we placed all of the emission at negative velocities on the SW side in the same quadrant (i.e., behind the star in Fig. 3A), with the emission from positive velocities placed in the opposite quadrant (i.e., in front of the star for Fig. 3A), and similarly for the NE side. Since the positive velocities on the SW side have no physical interpretation, the emission in the mostly empty quadrants in Fig. 3 indicate the level of noise in the deprojected images. Gas close to the line-of-sight (within 1 arcsec of the star) has a velocity too close to zero to allow the position to be determined, and so is masked in the deprojected images. Fig. 3 shows two possible deprojections of Fig. 2 with such a scheme, each of which has a corresponding physical interpretation.

Fig. 3A shows a distribution that closely matches that expected in a model in which a planet migrated out trapping planetesimals into its 2:1 and 3:2 resonances(*25*). In this interpretation, the bright SW clump, which arises from the superposition of material in these two resonances, must lie on the far side of the star, as this side of the disk is moving towards us, and the CO photodissociation 'tail' would trail behind the resonant clump. The clump orbits with the planet, but the CO produced in collisions in the clump moves at the Keplerian velocity of the resonance, and so leaves a tail that extends behind the clump at the synodic rate. In the case of a planet at 60AU this would be 900yr.

Fig. 3B shows the alternative explanation involving a single Mars-mass collision. The collision point is stationary and the CO 'tail' is due to gas released from material as it passes through the collision point(*25*). The rotation direction implies the orientation shown in Fig.3B, where the single clump lies on the near side of the star.

Alternative solutions are possible where the CO is distributed over both near and far sides; however, these are not favoured because the SW clump appears compact and it would require a rather symmetrical gas distribution on the near and far sides.

Determination of dust radial distribution from submm continuum data

The radial distributions of dust in Fig.3C were obtained by assuming there is a continuous distribution of dust in a broad axisymmetric disk. That distribution was modelled with 11 consecutive annuli, each of which characterized by its surface density; the sub-mm emission from dust in each annulus was calculated assuming black-body emissivities and temperatures for stellar luminosity 8.7L$_{sun}$ (*12*). The best fit to the surface density distribution was derived by iterating from the outermost annulus inwards, adjusting the density of the annulus under consideration to fit the disk surface brightness at a projected separation equal to the radius of the annulus. After all annuli have been modified once, the process is repeated a few times until the distribution converges. Since the dust distribution cannot be entirely axisymmetric, this procedure was performed on the two sides of the disk separately, obtaining the radial distributions on the SW and NE sides.